\documentclass[aps,pra,twocolumn,superscriptaddress,nobalancelastpage]{revtex4-1}
\usepackage{amsmath,amssymb,mathtools}
\usepackage[dvipdfmx]{graphicx}
\usepackage[dvipdfmx,colorlinks=true,bookmarks=true,citecolor=blue,linkcolor=blue,urlcolor=blue,breaklinks=true]{hyperref}

\graphicspath{{fig/}}
\usepackage{dcolumn}
\usepackage{bm}
\usepackage[T1]{fontenc}
\usepackage{textcomp}
\usepackage{color}
\usepackage{mathrsfs}

\begin{document}

\title{Magnonic thermal transport using the quantum Boltzmann equation
}

\author{Kouki Nakata}
\affiliation{Advanced Science Research Center, Japan Atomic Energy Agency, Tokai, Ibaraki 319-1195, Japan}

\author{Yuichi Ohnuma}
\affiliation{Research Center for Advanced Science and Technology (RCAST),
The University of Tokyo, Meguro-ku, Tokyo 153-8904, Japan
}

\date{\today}

\begin{abstract}
We present a formula for thermal transport in the bulk of Bose systems based on the quantum Boltzmann equation (QBE). First, starting from the quantum kinetic equation and using the Born approximation for impurity scattering, we derive the QBE of Bose systems and provide a formula for thermal transport subjected to a temperature gradient. Next, we apply the formula to magnons. Assuming a relaxation time approximation and focusing on the linear response regime, we show that the longitudinal thermal conductivity of the QBE exhibits the different behavior from the conventional. The thermal conductivity of the QBE reduces to the Drude-type in the limit of the quasiparticle approximation, while not in the absence of the approximation. Finally, applying the quasiparticle approximation to the QBE, we find that the correction to the conventional Boltzmann equation is integrated as the self-energy into the spectral function of the QBE, and this enhances the thermal conductivity. Thus we shed light on the thermal transport property of the QBE beyond the conventional.
\end{abstract}

\maketitle

\section{Introduction}
\label{sec:Sec.1}

The last decade has seen a rapid development of magnon-based spintronics, 
dubbed magnonics,
aiming at utilizing the quantized spin waves, magnons,
as a carrier of information~\cite{MagnonSpintronics}.
The main subject is the realization of efficient transmission of information using spins in insulating magnets.
For this purpose,
taking into account 
the fundamental difference
of the quantum-statistical properties between electrons and magnons,
i.e., fermions and bosons, respectively,
many magnonic analogues of electron transport 
have been established both theoretically and experimentally~\cite{ReviewMagnon},
with a particular focus on thermal transport,
e.g., the thermal Hall effect~\cite{onose,katsura,Matsumoto,Matsumoto2} 
and 
the Wiedemann-Franz law~\cite{magnonWF,KJD,KSJD,KSS}
for magnon transport.

The key ingredient in the study 
of thermal transport phenomena
is the Boltzmann equation~\cite{AMermin}.
As well as electron transport,
the conventional Boltzmann equation 
has been playing a central role in the study of magnon transport,
e.g., see Refs.~\cite{MagnonBoltzmann2014,MagnonBoltzmann2014-2,Basso,Basso2,Basso3,Okuma,MagnonBoltzmann2018,1/3noiseMagnon,MagnonBoltzmann2019,MagnonBoltzmannDrude};
the temperature gradient
${\mathbf{\nabla}} T$
drives a magnonic system out
of equilibrium.
The conventional Boltzmann equation describes the property of the system 
at time $t$ as
\begin{align}
\Big(&\frac{\partial }{\partial t}+
{\mathbf{v}}_{\mathbf{k}} \cdot  {\mathbf{\nabla}} T
\frac{\partial }{\partial T}\Big)
f_{{\mathbf{k}}, {\mathbf{r}}, t}    
=
I_{{\mathbf{k}}, {\mathbf{r}}, t} ,
\label{eqn:1} 
\end{align}
where
${\mathbf{v}}_{\mathbf{k}} :=
\partial \omega _{\mathbf{k}}/(\partial  {\mathbf{k}})
$
is the magnon velocity
for the energy dispersion relation
$\hbar \omega _{\mathbf{k}}$
in the wavenumber space 
${\mathbf{k}}$,
$\hbar$ represents the Planck constant,
$f_{{\mathbf{k}}, {\mathbf{r}}, t}  $
is the nonequilibrium Bose distribution function
of the absolute temperature $T$,
and
$ I_{{\mathbf{k}}, {\mathbf{r}}, t} $
is the collision integral
for a position ${\mathbf{r}}$.
Within the relaxation time approximation,
the longitudinal thermal conductivity of magnons 
in the bulk of magnets
is proportional to the relaxation time~\cite{KSJD,1/3noiseMagnon,MagnonBoltzmannDrude}.
Under some conditions,
the relaxation time coincides with the lifetime of magnons,
which is proportional to the inverse of the Gilbert damping constant $\alpha$~\cite{SeeAppendix}.
Thus the magnonic thermal conductivity
of the conventional Boltzmann equation
reduces to
the Drude-type~\cite{Drude-typeNote}
as a function of $\alpha$
in that it is proportional to $1/\alpha$.

From the viewpoint of quantum field theory,
the conventional Boltzmann equation [Eq.~\eqref{eqn:1}]
is derived from the quantum kinetic equation~\footnote{
Ref.~\cite{haug} refers to
the quantum kinetic equation
as
the Kadanoff-Baym equation~\cite{KadanoffBaym},
or 
the Keldysh equation~\cite{Keldysh}.
For the details,
see also Ref.~\cite{QuantumKineticEq1984}.
}
by taking several approximations~\cite{mahan,haug,kita,QuantumKineticEq1984}.
Assuming that
the variation of the center-of-mass coordinates 
is slow compared with that of the relative coordinates,
Kadanoff and Baym~\cite{KadanoffBaym} 
applied an approximation,
called the gradient expansion,
to the quantum kinetic equation for the nonequilibrium Green's function~\cite{Keldysh}.
The quantum kinetic equation
of the lowest order gradient approximation
becomes the quantum Boltzmann equation (QBE).
The QBE describes the equation of motion for
the lesser Green's function.
The spectral function is assumed to be the Dirac delta function
in the quasiparticle approximation~\cite{mahan,haug,kita,QPapproximation,1/3noiseMagnon}.
In the limit of the quasiparticle approximation,
the QBE reduces to the conventional Boltzmann equation [Eq.~\eqref{eqn:1}]
for the nonequilibrium distribution function
of three variables
$({\mathbf{k}},  {\mathbf{r}}, t)$.

This hierarchical structure of quantum field theory
indicates that
by relaxing some of the approximations,
quantum-mechanical corrections 
to the conventional Boltzmann equation can be evaluated~\cite{haug}.
Such sound development has been made successfully
as to electrons~\cite{kita}.
The lifetime of electrons in metals subjected to a strong impurity potential 
becomes substantial,
and the quasiparticle approximation is not applicable.
To solve the issue,
Prange and Kadanoff introduced an alternative approach~\cite{QCAkadanoff},
which was developed for the application to
superconductors and superfluids~\cite{QCAeilenberger,QCAsf,QCAlarkin,QCAkita},
e.g., the Eilenberger equation~\cite{QCAeilenberger}.
However, those are for Fermi systems.
Since the approach is based on the assumption that~\cite{kita}
there is a Fermi surface and the Fermi energy,
the developed formula is not applicable to Bose systems, 
e.g., magnons.
Thus, to the best of our knowledge~\cite{DanielPrivate},
as for magnons in the bulk of magnets,
the thermal transport property 
beyond the conventional Boltzmann equation 
remains an open issue.

In this paper, 
we provide a solution to this fundamental challenge
by 
starting from the quantum kinetic equation
and
developing the QBE for Bose systems.
The purpose of any useful formalism is to provide
a method for calculation of measurable quantities.
First, using the QBE we develop a formula for thermal transport 
in the bulk of Bose systems, 
including the nonlinear response to the temperature gradient.
Next, as a platform, we apply it to magnons.
In the conventional spintronics study, 
the Landau-Lifshitz-Gilbert equation is playing the central role~\cite{LLGspintroReview}. 
To develop a relation with it,
using the Gilbert damping constant we describe the spectral function of magnons,
and study the longitudinal thermal conductivity of the QBE.
Finally, by applying the quasiparticle approximation to the thermal conductivity of the QBE,
we find the correction to the conventional Boltzmann equation
and discuss thermal transport properties
beyond the conventional.

We remark that the conventional Boltzmann equation~\cite{MagnonBoltzmann2014,MagnonBoltzmann2014-2,Basso,Basso2,Basso3,Okuma,MagnonBoltzmann2018,1/3noiseMagnon,MagnonBoltzmann2019,MagnonBoltzmannDrude},
i.e., the transport theory based on the quasiparticle approximation,
can not describe paramagnons
in the bulk of paramagnets~\cite{mahan}.
The quasiparticle approximation assumes that the spectrum has the form of the Dirac delta function.
However, the spectrum of paramagnons is broad in general,
and has a peak with a sufficient width of a nonzero value
associated with
the inverse of the finite lifetime~\cite{Paramagnon_Doniach,paramagnonHertz,Paramagnon_observation2010};
the spectrum can not be approximated by the Dirac delta function.
Therefore, the conventional Boltzmann equation can not describe paramagnons.
In this paper, we will also shed light on this issue.

This paper is organized as follows.
In Sec.~\ref{sec:Sec.2}
starting from the quantum kinetic equation of the lowest order gradient approximation
and using the Born approximation for impurity scattering,
we derive the QBE for Bose systems.
In Sec.~\ref{sec:Sec.3},
first, using the QBE
and
assuming a steady state in terms of time,
we provide a formula for thermal transport 
in the bulk of Bose systems subjected to a temperature gradient,
including the nonlinear response.
Next, we apply the formula to magnons in Sec.~\ref{subsec:Sec.3-1}.
To develop a relation with the conventional spintronics study,
we describe the spectral function of magnons in terms of the Gilbert damping constant.
Then,
assuming a relaxation time approximation
and focusing on the linear response regime,
we evaluate the longitudinal thermal conductivity
of magnons in the bulk of magnets
based on the QBE.
Finally, in Sec.~\ref{subsec:Sec.3-2}
applying the quasiparticle approximation to the magnonic thermal conductivity of the QBE,
we discuss the difference from the one 
of the conventional Boltzmann equation.
Comparing also with the linear response theory,
we comment on our formula
in Sec.~\ref{sec:discussion}.
We remark on open issues in Sec.~\ref{sec:Outlook}
and give some conclusions in Sec.~\ref{sec:conclusion}.
Technical details are deferred to the Appendices.

\section{Quantum Boltzmann equation  
for Bose system}
\label{sec:Sec.2}

We consider a Bose system
where the center-of-mass coordinates,
the position and time in center-of-mass
(${\mathbf{r}}, t$), respectively,
vary slowly compared to the relative coordinates.
Up to the lowest order of the gradient expansion,
the quantum kinetic equation~\cite{mahan,haug,kita,QuantumKineticEq1984} for the system
reduces to
\begin{align}
 -i \Big(&
\frac{\partial {\mathscr{H}}_{{\mathbf{k}}, \omega}}{\partial  t} \frac{\partial }{\partial  \omega }
-\frac{\partial {\mathscr{H}}_{{\mathbf{k}}, \omega}}{\partial  \omega } \frac{\partial }{\partial  t} 
-\frac{\partial {\mathscr{H}}_{{\mathbf{k}}, \omega}}{\partial  {\mathbf{r}} } \frac{\partial }{\partial  {\mathbf{k}}}
+\frac{\partial {\mathscr{H}}_{{\mathbf{k}}, \omega}}{\partial  {\mathbf{k}} } \frac{\partial }{\partial  {\mathbf{r}}}
\Big)   G^{<}_{{\mathbf{k}}, \omega , {\mathbf{r}}, t}
\nonumber  \\
=& (G^{<}_{{\mathbf{k}}, \omega , {\mathbf{r}}, t} \Sigma ^{>}_{{\mathbf{k}}, \omega , {\mathbf{r}}, t}
- G^{>}_{{\mathbf{k}}, \omega , {\mathbf{r}}, t} \Sigma ^{<}_{{\mathbf{k}}, \omega , {\mathbf{r}}, t}), 
 \label{eqn:2} 
\end{align}
where
$  {\mathscr{H}}_{{\mathbf{k}}, \omega}:= \hbar  \omega  -  \hbar \omega _{\mathbf{k}} $
for a frequency $\omega$
and the functions,
$G^{<(>)}_{{\mathbf{k}}, \omega , {\mathbf{r}}, t}$
and
$\Sigma ^{<(>)}_{{\mathbf{k}}, \omega , {\mathbf{r}}, t}$,
are the lesser (greater) component of the bosonic nonequilibrium Green's function and that of the self-energy, respectively;
the variables 
$({\mathbf{k}}, \omega)$
arise from the Fourier transform of the relative coordinates. 
Following Ref.~\cite{mahan},
we refer to Eq.~\eqref{eqn:2}
as the QBE.
The QBE 
is the equation of motion for
the lesser Green's function
$G^{<}_{{\mathbf{k}}, \omega , {\mathbf{r}}, t}$
and consists of four variables
$({\mathbf{k}}, \omega , {\mathbf{r}}, t)$,
while the conventional Boltzmann equation 
is for the nonequilibrium Bose distribution function
$f_{{\mathbf{k}}, {\mathbf{r}}, t}  $
and of three variables
$({\mathbf{k}},  {\mathbf{r}}, t)$.
The QBE 
in the limit of the quasiparticle approximation
reduces to the conventional Boltzmann equation.
In this paper we study thermal transport of the QBE for Bose systems,
and find the properties beyond the conventional Boltzmann equation.
Then, applying the quasiparticle approximation to the thermal conductivity of the QBE,
we discuss the difference from the conventional.

Within the Born approximation,
the self-energy due to impurity scattering 
of the impurity potential
$  V_{{\mathbf{k}}, {\mathbf{k}}^{\prime}}$
is given as
$\Sigma _{{\mathbf{k}}, \omega , {\mathbf{r}}, t} 
= 
\sum_{{\mathbf{k}}^{\prime}} 
G_{{\mathbf{k}}^{\prime}, \omega , {\mathbf{r}}, t}
\mid  V_{{\mathbf{k}}, {\mathbf{k}}^{\prime}}   \mid ^2$.
Assuming that 
the function $ {\mathscr{H}}_{{\mathbf{k}}, \omega}$
is time-independent and spatially uniform,
the QBE becomes
\begin{align}
(&\partial_t+
{\mathbf{v}}_{\mathbf{k}} \cdot 
\partial_{\mathbf{r}})
G^{<}_{{\mathbf{k}}, \omega , {\mathbf{r}}, t}  
\label{eqn:3} \\
=
&\frac{1}{i \hbar}
\sum_{{\mathbf{k}}^{\prime}} 
\mid  V_{{\mathbf{k}}, {\mathbf{k}}^{\prime}}   \mid ^2   
(G^{<}_{{\mathbf{k}}, \omega , {\mathbf{r}}, t} 
G ^{>}_{{\mathbf{k}}^{\prime}, \omega , {\mathbf{r}}, t}
- G^{>}_{{\mathbf{k}}, \omega , {\mathbf{r}}, t} 
G^{<}_{{\mathbf{k}}^{\prime}, \omega , {\mathbf{r}}, t}).
\nonumber
\end{align}
The QBE consists of 
the lesser (greater) Green's functions
$G^{<(>)}_{{\mathbf{k}}, \omega , {\mathbf{r}}, t}$.
The Kadanoff-Baym ansatz ensures that~\cite{mahan,haug,kita}
the Green's functions are associated with 
the spectral function 
${\cal{A}}_{{\mathbf{k}}, \omega, {\mathbf{r}}, t}$
and
the nonequilibrium distribution function
$\phi_{{\mathbf{k}}, \omega, {\mathbf{r}}, t}$
as
$ G^{<}_{{\mathbf{k}}, \omega, {\mathbf{r}}, t}  
= -i  {\cal{A}}_{{\mathbf{k}}, \omega, {\mathbf{r}}, t} 
\phi_{{\mathbf{k}}, \omega, {\mathbf{r}}, t}  $
and
$ G^{>}_{{\mathbf{k}}, \omega, {\mathbf{r}}, t}  
= -i  {\cal{A}}_{{\mathbf{k}}, \omega, {\mathbf{r}}, t} 
(1+\phi_{{\mathbf{k}}, \omega, {\mathbf{r}}, t}) $
for bosons,
while
$ G^{<}_{{\mathbf{k}}, \omega, {\mathbf{r}}, t}  
= i  {\cal{A}}_{{\mathbf{k}}, \omega, {\mathbf{r}}, t} 
\phi_{{\mathbf{k}}, \omega, {\mathbf{r}}, t}  $
and
$ G^{>}_{{\mathbf{k}}, \omega, {\mathbf{r}}, t}  
= -i  {\cal{A}}_{{\mathbf{k}}, \omega, {\mathbf{r}}, t} 
(1-\phi_{{\mathbf{k}}, \omega, {\mathbf{r}}, t}) $ 
for fermions.
The nonequilibrium Bose distribution function 
in the wavenumber space is given as
$ f_{{\mathbf{k}}, {\mathbf{r}}, t}^{\textrm{QBE}} 
:=\int [\hbar d \omega/(2\pi)] 
{\cal{A}}_{{\mathbf{k}}, \omega, {\mathbf{r}}, t}
\phi_{{\mathbf{k}}, \omega , {\mathbf{r}}, t}  $.
Using the Kadanoff-Baym ansatz for bosons,
finally, we obtain the QBE of the functions
${\cal{A}}_{{\mathbf{k}}, \omega, {\mathbf{r}}, t}$
and
$\phi_{{\mathbf{k}}, \omega, {\mathbf{r}}, t}$
as
\begin{align}
(&\partial_t+
{\mathbf{v}}_{\mathbf{k}} \cdot 
\partial_{\mathbf{r}})
({\cal{A}}_{{\mathbf{k}}, \omega, {\mathbf{r}}, t} 
\phi_{{\mathbf{k}}, \omega, {\mathbf{r}}, t}) 
\label{eqn:4} \\
= - & \frac{1}{\hbar} \sum_{{\mathbf{k}}^{\prime}} 
\mid  V_{{\mathbf{k}}, {\mathbf{k}}^{\prime}}   \mid ^2
{\cal{A}}_{{\mathbf{k}}, \omega, {\mathbf{r}}, t} 
{\cal{A}}_{{\mathbf{k}}^{\prime}, \omega , {\mathbf{r}}, t } 
(\phi_{{\mathbf{k}}, \omega, {\mathbf{r}}, t} 
- \phi_{{\mathbf{k}}^{\prime}, \omega, {\mathbf{r}}, t }). \nonumber
\end{align}
The QBE is useful to a wide range of Bose system
subjected to impurity scattering.
For convenience, we define the collision integral as
${\cal{I}}_{{\mathbf{k}}, \omega, {\mathbf{r}}, t}
:= -  \sum_{{\mathbf{k}}^{\prime}} 
\mid  V_{{\mathbf{k}}, {\mathbf{k}}^{\prime}}   \mid ^2
{\cal{A}}_{{\mathbf{k}}, \omega, {\mathbf{r}}, t} 
{\cal{A}}_{{\mathbf{k}}^{\prime}, \omega , {\mathbf{r}}, t } 
(\phi_{{\mathbf{k}}, \omega, {\mathbf{r}}, t} 
- \phi_{{\mathbf{k}}^{\prime}, \omega, {\mathbf{r}}, t })/\hbar$.
Hereafter, for simplicity, 
we drop the indices
(${\mathbf{r}}, t$)
when those are not important.

\section{Thermal transport in \\
 Bose system}
\label{sec:Sec.3}

The temperature gradient drives the Bose system 
out of equilibrium
and generate a heat current.
The QBE [Eq.~\eqref{eqn:4}] describes
the transport property 
of a steady state in terms of time
as
\begin{align}
{\mathbf{v}}_{\mathbf{k}} \cdot  {\mathbf{\nabla}} T
\frac{\partial }{\partial T}
({\cal{A}}_{{\mathbf{k}}, \omega }
\phi_{{\mathbf{k}}, \omega}) 
={\cal{I}}_{{\mathbf{k}}, \omega},
\label{eqn:20} 
\end{align}
where
we assume that the temperature gradient is spatially uniform
${\mathbf{\nabla}} T=({\textrm{const.}})$.
In this section, first,
using the functions
$ {\cal{A}}_{{\mathbf{k}}, \omega} $
and
$\phi_{{\mathbf{k}}, \omega}$
of the QBE [Eq.~\eqref{eqn:20}],
we provide a formula
for the heat current 
in the bulk of two-dimensional Bose systems.
Next, as a platform, 
in Sec.~\ref{subsec:Sec.3-1}
we apply the formula to magnons 
in two-dimensional insulating magnets.
Assuming a relaxation time approximation 
for the function
${\cal{A}}_{{\mathbf{k}}, \omega}\phi_{{\mathbf{k}}, \omega} $
and
focusing on the linear response regime,
we evaluate the thermal conductivity 
in the bulk of the magnet.
Finally, in Sec.~\ref{subsec:Sec.3-2}
applying the quasiparticle approximation to the magnonic thermal conductivity of the QBE,
we discuss the difference from the one 
of the conventional Boltzmann equation.

The applied temperature gradient 
drives the system out of equilibrium
and 
the Bose distribution function 
$\phi_{{\mathbf{k}}, \omega}$
deviates from the one 
$\phi_0=({\text{e}}^{\beta \hbar \omega}-1)^{-1}$
in equilibrium,
where
$\beta:=1/(k_{\textrm{B}}T)$ is the inverse temperature
and 
$k_{\textrm{B}}$ represents the Boltzmann constant.
The deviation is characterized as the function
$ \delta  \phi_{{\mathbf{k}}, \omega} 
:= \phi_{{\mathbf{k}}, \omega} -\phi_0 
$.
Since the self-energy arises from impurity scattering,
we assume that the spectral function 
$ {\cal{A}}_{{\mathbf{k}}, \omega} $
is little influenced by temperature
and neglect the temperature dependence.
Therefore the nonequilibrium Bose distribution function 
in the wavenumber space is given as
$ f_{\mathbf{k}}^{\textrm{QBE}}  = 
f_0^{\textrm{QBE}} + \delta f_{\mathbf{k}}^{\textrm{QBE}}  $
with
$ f_0^{\textrm{QBE}} :=\int [\hbar d \omega/(2\pi)] {\cal{A}}_{{\mathbf{k}}, \omega}\phi_0  $
and
$ \delta f_{\mathbf{k}}^{\textrm{QBE}} :=\int [\hbar d \omega/(2\pi)] {\cal{A}}_{{\mathbf{k}}, \omega} \delta \phi_{{\mathbf{k}}, \omega} $.
The function consists of two parts;
the equilibrium component 
$ f_0^{\textrm{QBE}} $
and the nonequilibrium one
$ \delta f_{\mathbf{k}}^{\textrm{QBE}} $.
Since each mode $\omega$ 
subjected to a chemical potential $\mu$
carries the energy $\hbar \omega $,
the heat current density in the bulk of 
two-dimensional Bose systems,
${\mathbf{j}}_{Q}=(j_{Q_x}, j_{Q_y}) $,
is given as
\begin{align}
{\mathbf{j}}_{Q}=
\int \frac{d^2 {\mathbf{k}}}{(2\pi)^2}{\mathbf{v}}_{\mathbf{k}}
\int \frac{\hbar d \omega}{2\pi}
(\hbar \omega -\mu)
{\cal{A}}_{{\mathbf{k}}, \omega} \delta \phi_{{\mathbf{k}}, \omega}.
 \label{eqn:5} 
\end{align}
This is the formula for the heat current density of the QBE,
including the nonlinear response to the temperature gradient.
The formula [Eq.~\eqref{eqn:5}] is useful to Bose systems
(e.g., insulators and metals)
with the spectral function of arbitrary shape.

\subsection{Magnonic thermal conductivity}
\label{subsec:Sec.3-1}

As a platform, we apply the formula for the heat current of the QBE [Eq.~\eqref{eqn:5}]
to magnons in a two-dimensional insulating magnet
where time-reversal symmetry is broken,
e.g., due to an external magnetic field.
At sufficiently low temperatures,
the effect of magnon-magnon interactions and
that of phonons 
are negligibly small,
and
impurity scattering 
makes a major contribution to
the self-energy.
Therefore, we work under the assumption that the spectral function 
$ {\cal{A}}_{{\mathbf{k}}, \omega} $
is little influenced by temperature,
and neglect the temperature dependence.

First, we comment on the chemical potential of magnons
subjected to the temperature gradient
~\cite{magnonWF,KJD,KSJD}.
The applied temperature gradient induces magnon transport,
which leads to an accumulation of magnons at the boundaries
and builds up a nonuniform magnetization in the sample.
This magnetization gradient plays a role of 
an effective magnetic field gradient
and works as the gradient of a nonequilibrium spin chemical potential~\cite{Basso2,MagnonChemicalWees,YacobyChemical,demokritov} for magnons.
This generates a counter-current of magnons
and thus the nonequilibrium spin chemical potential 
contributes to the thermal conductivity.
See Ref.~\cite{KSJD} for details.
In this paper, for simplicity,
we consider a sufficiently large system
and work under the assumption that
the effect of the boundaries is negligibly small.
Consequently, the nonequilibrium magnon accumulation 
becomes negligible
and the nonequilibrium spin chemical potential of magnons vanishes.
In the magnonic system, 
the heat current is identified with the energy current~\cite{Matsumoto,KJD}.
Note that~\cite{Basso2,MagnonChemicalWees,YacobyChemical,demokritov} 
the spin chemical potential of magnons 
is peculiar to the system out of equilibrium~\footnote{
Ref.~\cite{Basso2} indicates that 
the nonequilibrium spin chemical potential
can be regarded as 
a Johnson-Silsbee potential~\cite{SilsbeeMagnetization}.}.

Then, we consider thermal transport carried by magnons
with the energy dispersion relation
$\hbar \omega _{\mathbf{k}} = Dk^2+\Delta$,
where 
$D$ represents the spin stiffness constant,
$k:=|{\mathbf{k}}|$
denotes the magnitude of the wavenumber,
and $\Delta$ is the magnon energy gap,
e.g., due to an external magnetic field and a spin anisotropy, etc.
Assuming a relaxation time approximation for the function
${\cal{A}}_{{\mathbf{k}}, \omega}\phi_{{\mathbf{k}}, \omega} $
of
\begin{align}
 {\cal{I}}_{{\mathbf{k}}, \omega}=
 -\frac{{\cal{A}}_{{\mathbf{k}}, \omega}\phi_{{\mathbf{k}}, \omega}
 -{\cal{A}}_{{\mathbf{k}}, \omega}\phi_0}{\tau_{{\mathbf{k}}, \omega}^{\textrm{R}}} 
 \label{eqn:6} 
\end{align}
and 
focusing on the linear response regime,
from Eq.~\eqref{eqn:20}
the nonequilibrium component 
$\delta \phi_{{\mathbf{k}}, \omega}$
is given as
\begin{align}
\delta \phi_{{\mathbf{k}}, \omega}
=-\tau_{{\mathbf{k}}, \omega}^{\textrm{R}}
{\mathbf{v}}_{\mathbf{k}} \cdot  {\mathbf{\nabla}} T 
\frac{\partial \phi_0}{\partial  T }+O(({\mathbf{\nabla}} T)^2),
 \label{eqn:7} 
\end{align}
where 
$\tau_{{\mathbf{k}}, \omega}^{\textrm{R}}$
is the relaxation time for the magnonic system.
Under the assumption that
impurity scattering is elastic 
and that
the relaxation time depends solely on
the magnitude of the wavenumber,
it is evaluated as
$
{1}/{\tau_{{\mathbf{k}}, \omega}^{\textrm{R}}}
= \sum_{{\mathbf{k}}^{\prime}} 
\mid  V_{{\mathbf{k}}, {\mathbf{k}}^{\prime}}   \mid ^2
 {\cal{A}}_{{\mathbf{k}}^{\prime}, \omega} 
(1-{{\mathbf{v}}_{\mathbf{k}}\cdot{\mathbf{v}}_{{\mathbf{k}^{\prime}}}}/{\mid   {\mathbf{v}}_{\mathbf{k}} \mid ^2})/\hbar
$.
We remark that the relaxation time 
$\tau_{{\mathbf{k}}, \omega}^{\textrm{R}}$
is different from the magnon lifetime
$\tau_{{\mathbf{k}}, \omega}^{\textrm{L}}$ in general.
Those are distinct quantities.
However,
when the impurity potential is localized in space,
the Fourier component $ V_{{\mathbf{k}}, {\mathbf{k}}^{\prime}}$
becomes independent of the wavenumber,
and the relaxation time coincides with the magnon lifetime,
which takes a wavenumber-independent value
as
$\tau_{ \omega}^{\textrm{R}}=\tau_{ \omega}^{\textrm{L}}$.
From the Landau-Lifshitz-Gilbert equation~\cite{LLGspintroReview},
the magnon lifetime is associated with
the inverse of the Gilbert damping constant $\alpha$
and it is described as~\cite{adachi,OhnumaSP,AGD,TataraMagnonLuttinger,KovalevEPL,MagnonBoltzmannDrude}
$\hbar/(2\tau_{ \omega}^{\textrm{L}})
= \alpha \hbar \omega$.
Therefore,
under the assumption that the real part of the self-energy is negligibly small compared with the magnon energy gap,
the spectral function is given as
${\cal{A}}_{{\mathbf{k}}, \omega}
=2 \alpha \hbar \omega/[({\mathscr{H}}_{{\mathbf{k}}, \omega})^2+(\alpha \hbar \omega)^2]$.
See Appendices for details~\cite{SeeAppendix}.

Finally, from Eqs.~\eqref{eqn:5} and~\eqref{eqn:7},
we obtain the longitudinal thermal conductivity of magnons
in the bulk of the two-dimensional insulating magnet,
$\kappa_{xx}:= -j_{Q_x}/(\partial _x T) $,
as
\begin{align}
\kappa_{xx}=&
\frac{1}{2}
\Big(\frac{D}{\pi}\Big)^2
\frac{1}{k_{\textrm{B}}T^2}
\int_0^{\infty} dk k^3  \nonumber \\
\cdot &
\int d \omega  
\frac{(\hbar \omega)^2}{({\mathscr{H}}_{{\mathbf{k}}, \omega})^2+(\alpha \hbar \omega)^2}
\frac{{\textrm{e}}^{\beta \hbar \omega}}{({\textrm{e}}^{\beta \hbar \omega}-1)^2},
 \label{eqn:8} 
\end{align}
where we assume that the temperature gradient is applied along the $x$ axis.
In contrast to the conventional Boltzmann equation
(cf., Sec.~\ref{sec:Sec.1}),
the thermal conductivity of the QBE 
in the absence of the quasiparticle approximation
does not reduce to the Drude-type~\cite{Drude-typeNote},
as a function of the Gilbert damping constant $\alpha$,
in that it is not proportional to $1/\alpha$.
The factor $1/\alpha$ arises from
the relaxation time 
$\tau_{ \omega}^{\textrm{R}}$
in
$\delta \phi_{{\mathbf{k}}, \omega}$
[Eq.~\eqref{eqn:7}]
as
$\tau_{ \omega}^{\textrm{R}}
=\tau_{\omega}^{\textrm{L}} 
={1}/(2 \alpha \omega)$.
However, it cancels out by
the factor $2 \alpha \hbar \omega$ 
of the spectral function
${\cal{A}}_{{\mathbf{k}}, \omega}
=2 \alpha \hbar \omega/[({\mathscr{H}}_{{\mathbf{k}}, \omega})^2+(\alpha \hbar \omega)^2]
$
[Eq.~\eqref{eqn:5}].
Therefore the integrand
remains the Lorentz-type~\footnote{
In this paper
we refer to the function of $\alpha$,
$F(\alpha):=C_1/({\alpha}^2+C_2)$ with
$C_1>0$ and $C_2>0$,
as a Lorentz-type
to distinguish from the Drude-type~\cite{Drude-typeNote}.
},
and the thermal conductivity of the QBE
does not reduce to the Drude-type
in the absence of the quasiparticle approximation.

\subsection{Comparison:
Conventional Boltzmann equation}
\label{subsec:Sec.3-2}

The QBE in the limit of the quasiparticle approximation
reduces to the conventional Boltzmann equation,
which provides the thermal conductivity of the Drude-type~\cite{Drude-typeNote}
in terms of the Gilbert damping constant $\alpha$
(cf., Sec.~\ref{sec:Sec.1}).
This agrees with our formula of the QBE 
[Eq.~\eqref{eqn:5}];
when we employ the quasiparticle (qp) approximation
${\cal{A}}_{{\mathbf{k}}, \omega }
\approx
{\cal{A}}_{{\mathbf{k}}, \omega}^{\textrm{qp}}
:= 2 \pi \delta ( {\mathscr{H}}_{{\mathbf{k}}, \omega})$
for Eq.~\eqref{eqn:5},
the thermal conductivity of the QBE
reduces to
$\kappa_{xx} \approx \kappa_{xx}^{\textrm{qp}} $
as
\begin{align}
\kappa_{xx}^{\textrm{qp}}
=
\frac{1}{2}
\Big(\frac{D}{\pi}\Big)^2
\frac{1}{k_{\textrm{B}}T^2}
\frac{\pi}{\alpha}
\int_0^{\infty} dk k^3 \omega _{\mathbf{k}}
\frac{{\textrm{e}}^{\beta \hbar \omega _{\mathbf{k}}}}{({\textrm{e}}^{\beta \hbar \omega _{\mathbf{k}}}-1)^2}.
 \label{eqn:9} 
\end{align}
This is consistent with Eq.~\eqref{eqn:8} in the limit of
$\alpha  \rightarrow  0$.
Thus we find in the limit of the quasiparticle approximation
that the thermal conductivity of the QBE becomes the Drude-type, 
as a function of $\alpha$,
in that it is proportional to $1/\alpha$
as
$\kappa_{xx} \approx \kappa_{xx}^{\textrm{qp}} 
\propto  {1}/{\alpha}$.
The factor $1/\alpha$ arises from
the relaxation time
$\tau_{ \omega}^{\textrm{R}}
=\tau_{\omega}^{\textrm{L}} 
={1}/(2 \alpha \omega)$
in
$\delta \phi_{{\mathbf{k}}, \omega}$
[Eq.~\eqref{eqn:7}].

\begin{figure}[t]
\begin{center}
\includegraphics[width=8.8cm,clip]{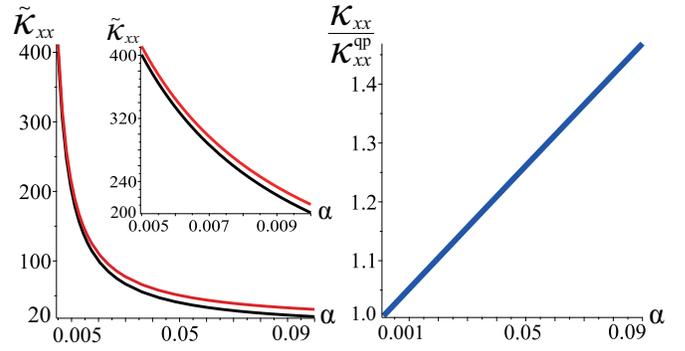}
\caption{
(Left)
Plots of the rescaled thermal conductivity of magnons,
${\tilde{\kappa}}_{xx}
:=[2\pi^2\hbar/(k_{\textrm{B}}^2T)]
{\kappa}_{xx}$,
as a function of the Gilbert damping constant 
$\alpha$
~\cite{GilbertInsulator,GilbertMetal}
obtained by numerically solving 
Eqs.~\eqref{eqn:8} and ~\eqref{eqn:9}
for $\Delta = k_{\textrm{B}}T$
with a fixed temperature.
Red line:
Eq.~\eqref{eqn:8} of the QBE.
Black line:
Eq.~\eqref{eqn:9}
for 
$\kappa_{xx} \approx \kappa_{xx}^{\textrm{qp}} $
in the limit of the quasiparticle approximation,
i.e., the conventional Boltzmann equation.
The correction to the conventional Boltzmann equation
enhances the thermal conductivity.
Inset:
The enhancement still works and remains significant 
even for small values of the parameter $\alpha$.
(Right)
The plot of the ratio, 
$\kappa_{xx}/\kappa_{xx}^{\textrm{qp}} 
=1+(\kappa_{xx} - \kappa_{xx}^{\textrm{qp}} )/\kappa_{xx}^{\textrm{qp}}$,
as a function of the Gilbert damping constant $\alpha$,
i.e., the ratio of the magnonic thermal conductivity
$\kappa_{xx}$
of the QBE
to the one
$\kappa_{xx}^{\textrm{qp}}$
of the conventional Boltzmann equation
(i.e., the QBE in the limit of the quasiparticle approximation).
In units of 
$\kappa_{xx}^{\textrm{qp}}$,
the correction to the conventional one
and 
the resulting enhancement of the magnonic thermal conductivity
increase as the value of $\alpha$ becomes large.
}
\label{fig:QBE}
\end{center}
\end{figure}

In conclusion,
using the QBE we have developed the formula for thermal transport in the bulk of Bose systems,
with a particular focus on magnons.
As a function of the Gilbert damping constant,
the thermal conductivity of the QBE 
reduces to the Drude-type~\cite{Drude-typeNote}
in the limit of the quasiparticle approximation,
while not in the absence of the approximation.
This is the main difference in the thermal conductivity
between
the QBE
and the conventional one.

We remark that
our formula based on the QBE
reduces to the Drude-type~\cite{Drude-typeNote}
in the limit of the quasiparticle approximation.
This means that,
by relaxing the quasiparticle approximation,
the correction to the conventional Boltzmann equation 
is integrated as the self-energy
into the spectral function 
${\cal{A}}_{{\mathbf{k}}, \omega, {\mathbf{r}}, t}$
of the QBE;
Fig.~\ref{fig:QBE} shows that 
the correction to the conventional Boltzmann equation
enhances the thermal conductivity of the QBE.
Thus our thermal transport theory
of the QBE with four variables
$({\mathbf{k}}, \omega , {\mathbf{r}}, t)$
is identified as an appropriate extension of 
the one of 
the conventional Boltzmann equation
with three variables $({\mathbf{k}},  {\mathbf{r}}, t)$.

\section{Discussion}
\label{sec:discussion}

To conclude a few comments on our approach are in order.
First,
the conventional Boltzmann equation~\cite{MagnonBoltzmann2014,MagnonBoltzmann2014-2,Basso,Basso2,Basso3,Okuma,MagnonBoltzmann2018,1/3noiseMagnon,MagnonBoltzmann2019,MagnonBoltzmannDrude}
[Eq.~\eqref{eqn:1}],
i.e., the transport theory based on the quasiparticle approximation, can not describe paramagnons~\cite{mahan}.
The quasiparticle approximation assumes that the spectrum has the form of the Dirac delta function.
However, the spectrum of paramagnons is broad in general,
and has a peak with a sufficient width of a nonzero value
associated with
the inverse of the finite lifetime~\cite{Paramagnon_Doniach,paramagnonHertz,Paramagnon_observation2010};
the spectrum of paramagnons can not be approximated by the Dirac delta function.
Therefore, the conventional Boltzmann equation can not describe paramagnons.
On the other hand,
our formula based on the QBE is applicable to magnons 
with the spectrum of arbitrary shape [Eq.~\eqref{eqn:5}].
In that sense, it is expected that
our thermal transport theory is useful also to paramagnons in the bulk of paramagnets~\footnote{
As for spin transport in paramagnets,
see Refs.~\cite{ParaSP,ParaSSE,ParamagnonTransport,ParaOkamoto}.
}.

We remark that
as well as insulators,
our formula [Eq.~\eqref{eqn:5}] is applicable,
in principle, also to metals;
the spectral function is described 
in terms of the Gilbert damping constant~\cite{SeeAppendix},
and the value for metals
(e.g., transition metal ferromagnets)~\cite{TransitionMetalFM}
$\alpha=O(10^{-2})$
is large compared with that for insulators 
$\alpha=O(10^{-3})$
in general~\cite{LLGspintroReview,GilbertInsulator,GilbertMetal}.
Fig.~\ref{fig:QBE} 
shows the behavior of the magnonic thermal conductivity
in the region
$O(10^{-3})\leq  \alpha \leq O(10^{-2})$.
As seen in the right of Fig.~\ref{fig:QBE},
$\kappa_{xx}/\kappa_{xx}^{\textrm{qp}} 
=1+(\kappa_{xx} - \kappa_{xx}^{\textrm{qp}} )/\kappa_{xx}^{\textrm{qp}}$,
in units of 
$\kappa_{xx}^{\textrm{qp}}$
the correction to the conventional Boltzmann equation
and 
the resulting enhancement of the magnonic thermal conductivity
increase as the value of $\alpha$ becomes large.
The damping constant is associated with
the inverse of the magnon lifetime~\cite{SeeAppendix}.

Next,
our formula based on the QBE has an advantage 
over the linear response theory that
it includes the nonlinear response;
Eq.~\eqref{eqn:5} is the formula for the heat current density
including the nonlinear response to the temperature gradient.
Here, using the QBE
we develop an analysis on the nonlinear response.
We apply the relaxation time approximation [Eq.~\eqref{eqn:6}] 
for the function 
${\cal{A}}_{{\mathbf{k}}, \omega}\phi_{{\mathbf{k}}, \omega} $
to the QBE [Eq.~\eqref{eqn:20}].
Since the relaxation is induced by impurity scattering,
we assume that the relaxation time is little influenced by temperature,
and neglect the temperature dependence.
Using the method of successive substitution,
the nonequilibrium component of the Bose distribution function 
$\delta \phi_{{\mathbf{k}}, \omega}$
[Eq.~\eqref{eqn:5}]
is evaluated,
beyond the liner response regime,
as
$\delta \phi_{{\mathbf{k}}, \omega}
=\sum_{n=1}^{\infty}
(-\tau_{{\mathbf{k}}, \omega}^{\textrm{R}}
{\mathbf{v}}_{\mathbf{k}} \cdot  {\mathbf{\nabla}} T)^n 
[{\partial^n \phi_0}/({\partial  T^n })]$,
where
the nonequilibrium component
$\delta \phi_{{\mathbf{k}}, \omega}$
is arranged in terms of the function
$ ({\mathbf{\nabla}} T)^n $.
Lastly,
combining
this equation
with
Eq.~\eqref{eqn:5},
we can obtain each coefficient of the nonlinear response
to the temperature gradient.
The formula is not restricted to magnonic systems,
and it is useful to a wide range of Bose systems.

Finally,
we add a comment to the linear response theory.
According to Ref.~\cite{TataraMagnonLuttinger},
which focuses on the region 
$\alpha  \ll  1$,
the linear response theory (i.e., Kubo formula)
results in the magnonic thermal conductivity of the Drude-type~\cite{Drude-typeNote},
the same as the conventional Boltzmann equation,
in that it is proportional to the inverse of the Gilbert damping constant.
Note that
the temperature gradient is not mechanical force
but thermodynamic force;
the temperature gradient
is not described by a microscopic Hamiltonian.
Therefore it is not straightforward to integrate the temperature gradient into the linear response theory.
Ref.~\cite{TataraMagnonLuttinger} described the effect of the temperature gradient with the help of the thermal vector potential ~\cite{TataraLuttinger} associated with the Luttinger's approach~\cite{Luttinger1964}.
For the details, see Ref.~\cite{TataraMagnonLuttinger}.
We remark that the Boltzmann equation has no difficulty in integrating the temperature gradient into the formalism, 
e.g., see Eq.~\eqref{eqn:1}.

\section{Outlook}
\label{sec:Outlook}

We give a few perspectives on further research.
As a platform,
in Sec.~\ref{subsec:Sec.3-1}
and Sec.~\ref{subsec:Sec.3-2}
focusing on 
insulating magnets
at sufficiently low temperature,
we have studied thermal transport of magnons
under the assumption that
the effect of magnon-magnon interactions and
that of phonons
are negligibly small,
and that elastic impurity scattering 
makes a major contribution to
the self-energy.
It is of interest to study those effects on magnonic thermal transport of the QBE,
with considering also the case that impurity scattering is inelastic~\footnote{It is expected from Ref.~\cite{ArakawaWL} that the weak localization of magnons is induced
in a disordered magnonic system
where impurities are randomly distributed
and time-reversal symmetry holds effectively.}.

In this paper
relaxing the quasiparticle approximation
and
using the QBE for Bose systems,
we have found that
the correction to the conventional Boltzmann equation 
enhances the thermal conductivity.
Therefore, based on the QBE
it is intriguing to study the effect of the correction
on the magnonic Wiedemann-Franz law~\cite{magnonWF,KJD,KSJD,KSS};
at low temperatures magnon transport
obeys a magnonic analogue of the Wiedemann-Franz law~\cite{WFgermany},
a universal law,
in that the ratio of heat to spin conductivity is linear in temperature and does not depend on material parameters 
except the $g$-factor. 
Note that
the magnonic Wiedemann-Franz law for the bulk of magnets
has been proposed 
in Ref.~\cite{KSJD}
based on the conventional Boltzmann equation.
For the magnonic Wiedemann-Franz law of the QBE,
the spin conductivity 
and the off-diagonal elements of the Onsager coefficient~\cite{KSJD}
remain to be obtained.
We believe it can be evaluated
by following the study~\cite{mahan}
on the electrical conductivity of the QBE
and developing it into the Bose system.
Using the QBE,
it will be intriguing also to study the magnonic Hall coefficients
in topologically nontrivial magnonic systems~\cite{katsura,Matsumoto,Matsumoto2,KJD,KSJD,KSS}.

\section{Conclusion}
\label{sec:conclusion}

Developing the quantum Boltzmann equation,
we have provided the formula for thermal transport
in the bulk of Bose systems,
including the nonlinear response to the temperature gradient.
We have then applied the formula to magnons
and have shown that thermal transport of the quantum Boltzmann equation exhibits the different behavior 
from the conventional.
The longitudinal thermal conductivity 
of the quantum Boltzmann equation
reduces to the Drude-type 
in the limit of the quasiparticle approximation,
while not in the absence of the approximation.
Relaxing the quasiparticle approximation,
we have found that
the correction to the conventional Boltzmann equation 
is integrated as the self-energy into the spectral function of the quantum Boltzmann equation, 
and this enhances the thermal conductivity. 
Our formula
is useful to Bose systems,
including metals as well as insulators,
with the spectral function of arbitrary shape.
Using
the quantum Boltzmann equation
we have found
the thermal transport property
beyond the conventional.

\acknowledgements

We would like to thank D. Loss
for 
the collaborative work on the related study
and
for
turning our attention to this subject;
this work is motivated by the discussion 
at Basel in 2015 (K.N.).
We are grateful also to 
T. Kita 
for 
educating the author (K.N.) 
on the theoretical method of this study
and
creating an opportunity to work on this subject,
and to Y. Araki and H. Chudo for helpful feedback.
We acknowledge support
by JSPS KAKENHI Grant Number JP20K14420 (K. N.), 
by Leading Initiative for Excellent Young Researchers, MEXT, Japan (K. N.),
and 
by JST ERATO Grant No. JPMJER1601(Y. O.).

\appendix
\renewcommand\thefigure{\thesection.\arabic{figure}}

\section{Relaxation time and magnon lifetime}
\label{sec:Rtime}

In this Appendix,
we show that 
the relaxation time coincides with the magnon lifetime
and takes a wavenumber-independent value
under the assumption that;
the relaxation time depends solely on
the magnitude of the wavenumber,
impurity scattering is elastic, 
and the impurity potential is localized in space.
We remark that
at sufficiently low temperatures,
the effect of magnon-magnon interactions and
that of phonons
are negligibly small,
and
impurity scattering 
makes a major contribution to
the self-energy.
Therefore, we assume that the spectral function 
is little influenced by temperature
and neglect the temperature dependence.

We consider magnons
with the energy dispersion relation of
$\hbar \omega _{\mathbf{k}} = Dk^2+\Delta$,
where 
$|{\mathbf{k}}|=:k$
denotes
the magnitude of the wavenumber.
First, 
assuming the steady state in terms of time
and
applying the relaxation time approximation for the function
${\cal{A}}_{{\mathbf{k}}, \omega}\phi_{{\mathbf{k}}, \omega} $
to the QBE (see the main text),
within the linear response regime
we obtain the nonequilibrium component
$\delta \phi_{{\mathbf{k}}, \omega}$
as
$\delta \phi_{{\mathbf{k}}, \omega}
=-\tau_{{\mathbf{k}}, \omega}^{\textrm{R}}
{\mathbf{v}}_{\mathbf{k}} \cdot  {\mathbf{\nabla}} T 
[\partial \phi_0 /(\partial  T) ]
+O(({\mathbf{\nabla}} T)^2)  $.
Next,
using the relaxation time approximation
for the collision integral
${\cal{I}}_{{\mathbf{k}}, \omega}
= -  \sum_{{\mathbf{k}}^{\prime}} 
\mid  V_{{\mathbf{k}}, {\mathbf{k}}^{\prime}}   \mid ^2
{\cal{A}}_{{\mathbf{k}}, \omega} 
{\cal{A}}_{{\mathbf{k}}^{\prime}, \omega } 
(\phi_{{\mathbf{k}}, \omega} 
- \phi_{{\mathbf{k}}^{\prime}, \omega })/\hbar$,
we reach
$\delta \phi_{{\mathbf{k}}, \omega}/\tau_{{\mathbf{k}}, \omega}^{\textrm{R}}
= \sum_{{\mathbf{k}}^{\prime}} 
\mid  V_{{\mathbf{k}}, {\mathbf{k}}^{\prime}}   \mid ^2
{\cal{A}}_{{\mathbf{k}}^{\prime}, \omega } 
(\delta\phi_{{\mathbf{k}}, \omega} 
- \delta \phi_{{\mathbf{k}}^{\prime}, \omega })/\hbar$.
Finally, combining the equations
under the assumption that
the relaxation time depends solely on
the magnitude of the wavenumber
and that
impurity scattering is elastic,
we obtain the relaxation time as
${1}/{\tau_{{\mathbf{k}}, \omega}^{\textrm{R}}}
= \sum_{{\mathbf{k}}^{\prime}} 
\mid  V_{{\mathbf{k}}, {\mathbf{k}}^{\prime}}   \mid ^2
 {\cal{A}}_{{\mathbf{k}}^{\prime}, \omega} 
(1-{{\mathbf{v}}_{\mathbf{k}}\cdot{\mathbf{v}}_{{\mathbf{k}^{\prime}}}}/{\mid   {\mathbf{v}}_{\mathbf{k}} \mid ^2})/\hbar$.

Since we assume that impurities are dilute,
the effect can be taken into account
within the Born approximation,
see the main text for details.
When the impurity potential is localized in space,
the Fourier component becomes independent of the wavenumber
and
it is described as
$ |V_{{\mathbf{k}}, {\mathbf{k}}^{\prime}}|^2=:
u^2 n_{\textrm{imp}}$,
where
$n_{\textrm{imp}}$ is the impurity concentration~\cite{haug}.
Then,
the self-energy 
becomes independent of the wavenumber ${\mathbf{k}}$,
and it is given as
$\Sigma _{{\mathbf{k}}, \omega }
=
\Sigma _{ \omega} 
:= u^2 n_{\textrm{imp}}
\sum_{{\mathbf{k}}^{\prime}} 
G_{{\mathbf{k}}^{\prime}, \omega }$.
Therefore, the spectral function 
${\cal{A}}_{{\mathbf{k}}, \omega}$
depends solely on the magnitude of the wavenumber 
$|{\mathbf{k}}|=:k$
and it is denoted as 
${\cal{A}}_{{\mathbf{k}}, \omega}
={\cal{A}}_{k, \omega}$.
We remark that 
the spectral function 
${\cal{A}}_{{\mathbf{k}}, \omega}$
consists of the self-energy
$\Sigma _{{\mathbf{k}}, \omega }
=
\Sigma _{ \omega} $
and the function 
${\mathscr{H}}_{{\mathbf{k}}, \omega}
={\mathscr{H}}_{k, \omega}$;
since we assume that 
the energy dispersion relation of magnons 
takes the form of
$\hbar \omega _{\mathbf{k}} = Dk^2+\Delta$,
the function
$  {\mathscr{H}}_{{\mathbf{k}}, \omega}:= \hbar  \omega  -  \hbar \omega _{\mathbf{k}} $
depends only on the magnitude of the wavenumber 
and it is represented as
${\mathscr{H}}_{{\mathbf{k}}, \omega}
={\mathscr{H}}_{k, \omega}$.
Thus, the spectral function 
becomes dependent only on the magnitude of the wavenumber
as
${\cal{A}}_{{\mathbf{k}}, \omega}
={\cal{A}}_{k, \omega}$.
Using this result with the relation
$ \int_0^{2\pi} d\theta {\textrm{cos}}{\theta}=0  $,
finally,
we obtain the relaxation time as
$ {1}/{\tau_{{\mathbf{k}}, \omega}^{\textrm{R}}}
= {1}/{\tau_{\omega}^{\textrm{R}}}
:=
u^2 n_{\textrm{imp}}
\sum_{{\mathbf{k}}^{\prime}} 
 {\cal{A}}_{{\mathbf{k}}^{\prime}, \omega}/\hbar$,
and find that it is independent of the wavenumber 
$ {\mathbf{k}} $.

The relaxation time coincides with the lifetime
for the impurity potential of
$ |V_{{\mathbf{k}}, {\mathbf{k}}^{\prime}}|^2=
u^2 n_{\textrm{imp}}$.
The lifetime
${\tau_{{\mathbf{k}}, \omega}^{\textrm{L}}}$
is associated with the imaginary part of the self-energy 
and it is described as
$\hbar/(2 {\tau_{{\mathbf{k}}, \omega}^{\textrm{L}}})
:=- {\textrm{Im}}\Sigma _{{\mathbf{k}}, \omega}^{\textrm{r}}  $
in general,
where
$\Sigma _{{\mathbf{k}}, \omega}^{\textrm{r}} $
represents the retarded component of the self-energy
and it is given as
$\Sigma _{{\mathbf{k}} \omega}^{\textrm{r}}
=\Sigma _{ \omega}^{\textrm{r}}
:= u^2 n_{\textrm{imp}}
\sum_{{\mathbf{k}}^{\prime}} 
G_{{\mathbf{k}}^{\prime}, \omega }^{\textrm{r}}$
for the impurity potential of
$ |V_{{\mathbf{k}}, {\mathbf{k}}^{\prime}}|^2=
u^2 n_{\textrm{imp}}$.
Since the imaginary part of the retarded Green's function is associated with the spectral function as
${\textrm{Im}}
G_{{\mathbf{k}}^{\prime}, \omega }^{\textrm{r}}
=- {\cal{A}}_{{\mathbf{k}}^{\prime}, \omega}/2$,
the lifetime becomes 
$ {1}/{\tau_{{\mathbf{k}}, \omega}^{\textrm{L}}}
= {1}/{\tau_{\omega}^{\textrm{L}}}
:=
u^2 n_{\textrm{imp}}
\sum_{{\mathbf{k}}^{\prime}} 
 {\cal{A}}_{{\mathbf{k}}^{\prime}, \omega}/\hbar$,
and takes the wavenumber-independent value.
Thus the lifetime coincides with
the relaxation time
${1}/{\tau_{\omega}^{\textrm{L}}}
={1}/{\tau_{\omega}^{\textrm{R}}}$.
 
We stress that the relaxation time 
is different from the lifetime in general.
Those are distinct quantities.
However,
under the assumption that
the relaxation time depends solely on
the magnitude of the wavenumber,
impurity scattering is elastic, 
and the impurity potential is localized in space,
the relaxation time coincides with the lifetime
and takes the wavenumber-independent value.

\section{Magnon spectral function and \\
\  \   \  \  \  \  \  \   \   \   \  \  \   
Gilbert damping constant}
\label{sec:Rtime}

In this Appendix,
we describe the spectral function of magnons
in terms of the Gilbert damping constant $\alpha$.
The Landau-Lifshitz-Gilbert equation
is playing the central role
in the conventional spintronics study~\cite{LLGspintroReview}.
Therefore to develop a relation with it,
it is useful to describe the spectrum
as a function of the Gilbert damping constant.
First, as seen above,
the spectral function 
${\cal{A}}_{{\mathbf{k}}, \omega}
={\cal{A}}_{k, \omega}$
consists of
the function
${\mathscr{H}}_{{\mathbf{k}}, \omega}
={\mathscr{H}}_{k, \omega}$
and
the self-energy
$\Sigma _{{\mathbf{k}}, \omega}
=\Sigma _{ \omega} $,
and it is described as~\cite{mahan,haug,kita}
${\cal{A}}_{k, \omega}
= -2 {\textrm{Im}} \Sigma _{ \omega}^{\textrm{r}}/
[({\mathscr{H}}_{k, \omega})^2
+({\textrm{Im}} \Sigma _{ \omega}^{\textrm{r}})^2]$,
where we assume that the real part of the self-energy is negligibly small compared with the magnon energy gap.
Next, 
the lifetime is defined as
the imaginary part of the self-energy in general as
$\hbar/(2 {\tau_{ \omega}^{\textrm{L}}})
:= - {\textrm{Im}}\Sigma _{ \omega}^{\textrm{r}}$.
Since the lifetime of magnons
is associated with the inverse of the Gilbert damping constant
as~\cite{adachi,OhnumaSP,AGD,TataraMagnonLuttinger,KovalevEPL,MagnonBoltzmannDrude}
$\hbar/(2 {\tau_{ \omega}^{\textrm{L}}})
= \alpha \hbar \omega$,
the imaginary part of the self-energy
is characterized in terms of the Gilbert damping constant.
Finally,
the spectral function 
is described as a function of the Gilbert damping constant as
${\cal{A}}_{k, \omega}
=2 \alpha \hbar \omega/[({\mathscr{H}}_{k, \omega})^2+(\alpha \hbar \omega)^2]$.

\bibliography{PumpingRef}

\begin{thebibliography}{65}
\bibitem [{Dru()}]{Drude-typeNote}%
  \BibitemOpen
  \href@noop {} {}\bibinfo {note} {The conductivity of the Drude model is
  proportional to the relaxation time~\cite{AMermin}. When the relaxation time
  coincides with the lifetime, it becomes proportional to the lifetime and we
  refer to the conductivity as the Drude-type. Since the magnon lifetime is
  proportional to the inverse of the Gilbert damping
  constant~\cite{adachi,OhnumaSP,AGD,TataraMagnonLuttinger,KovalevEPL,MagnonBoltzmannDrude},
  the magnon conductivity being proportional to $1/\alpha$ can be classified
  into a Drude-type for magnon transport. Thus, throughout this paper, we refer
  to the magnon conductivity being proportional to $1/\alpha$ as the Drude-type
  for convenience.}\BibitemShut {Stop}%
\end{thebibliography}
\end{document}